\documentclass[12pt]{article}
\pdfoutput=1

\usepackage{amsmath}
\usepackage{amssymb}

\usepackage{graphicx}
\usepackage{cite}
\usepackage{multirow}
\usepackage{longtable}
\usepackage{lscape}
\usepackage{lineno}

\allowdisplaybreaks[4] 

\makeatletter
\renewcommand{\fnum@table}{\textbf{\tablename~\thetable}}
\renewcommand{\fnum@figure}{\textbf{\figurename~\thefigure}}
\makeatother

\newcommand{\bi}{\begin{itemize}}
\newcommand{\ei}{\end{itemize}}

\newcommand{\be}{\begin{equation}}
\newcommand{\ee}{\end{equation}}
\newcommand{\bea}{\begin{eqnarray}}
\newcommand{\eea}{\end{eqnarray}}

\begin{document}

\renewcommand{\thefootnote}{\alph{footnote}}

\vspace*{-3.cm}
\begin{flushright}
FERMILAB-CONF-12-420-APC
\end{flushright}

\vspace*{0.3cm}

\renewcommand{\thefootnote}{\fnsymbol{footnote}}
\setcounter{footnote}{-1}

{\begin{center}
{\Large\bf
A Staged Muon-Based Neutrino and\\[.05in] Collider Physics Program}
 \end{center}}
\renewcommand{\thefootnote}{\alph{footnote}}

 \vspace{-.2cm}
\begin{minipage}[b]{0.33\linewidth}
\rightline{\includegraphics[width=2cm]{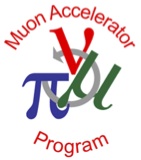}~~~\vspace{-.25cm}}
\end{minipage}
\begin{minipage}[b]{0.66\linewidth}
\normalsize
D. M. Kaplan (ed.), Illinois Institute of Technology,\\ for the Muon Accelerator Program (MAP) Collaboration
            
{\normalsize \vspace{0.20cm} 
July 31, 2012
}\end{minipage} 	 

\vspace*{0.5cm}

\renewcommand{\thefootnote}{\arabic{footnote}}
\setcounter{footnote}{0}

We sketch a staged plan for a series of muon-based facilities that can do compelling physics at each stage. Such a plan is unique in its ability to span both the Intensity and Energy Frontiers as defined by the P5 sub-panel of the US High Energy Physics Advisory Committee~\cite{P5-report}. This unique physics 
reach places a muon-based facility in an unequaled position to address critical questions about the 
nature of the Universe.   An R\&D program in support of the plan is in progress under the auspices of the US Muon Accelerator Program~\cite{MAP}. 
The plan is conceived in four stages, whose exact order remains to be worked out:
\begin{itemize}

\item The ``entry point" for the plan is the $\nu$STORM facility proposed at Fermilab, which can advance short-baseline physics by making definitive observations or exclusions of sterile neutrinos. 
Secondly, it can make key measurements to reduce systematic uncertainties in long-baseline neutrino experiments.  Finally, it can serve as an R\&D platform for demonstration of accelerator capabilities pre-requisite to the later stages.

\item A stored-muon-beam Neutrino Factory can take advantage of the large value of $\theta_{13}$ recently measured in reactor-antineutrino experiments to make definitive measurements of neutrino oscillations and their possible violation of CP symmetry.

\item Thanks to suppression of radiative effects by the muon mass and the $m_{\rm lepton}^2$ proportionality of the $s$-channel Higgs coupling, a ``Higgs Factory" Muon Collider can make uniquely precise measurements of the 126\,GeV boson recently discovered at the LHC. 

\item An energy-frontier Muon Collider can perform unique measurements of Terascale physics, offering both precision and discovery reach.

\end{itemize}
Stored-muon-beam facilities have been shown to be feasible in a number of studies. They exploit the relatively long muon lifetime of 2.2\,$\mu$s together with relativistic time dilation and rapid acceleration technologies. These advantages make muons the ``lepton of choice" for a variety of physics goals. The plan thus has the right ``footprint" to be the next major arena for High Energy Physics, complementing the LHC, over the next two decades.

In the following we first summarize the physics landscape to which muon facilities can contribute. We then briefly describe the main features of stored-muon-beam facilities. They present unparalleled near- and far-term opportunities that ideally will be exploited through cooperation among the regions of the High Energy Physics world.

\section{Short-Baseline Neutrino Physics}

In the present context, short-baseline physics refers to those neutrino flavor conversion/dis\-appearance phenomena which take place at $L/E$ values  considerably smaller than those
typically associated with the mass-squared splittings of atmospheric
and solar neutrino oscillations. In the last 18 months, this area has
seen a greatly increased scientific interest, resulting in a number
of workshops and documents, notably the sterile neutrino white paper~\cite{wp} 
and the report of the short-baseline focus group at Fermilab\cite{Brice:2012zz}. Therefore, it is
sufficient to summarize here the hints for oscillations at $L/E\sim
1\,\mathrm{m}/\mathrm{MeV}$ and refer the reader to the
aforementioned reports for more details.

There are the LSND~\cite{LSND} and now MiniBooNE~\cite{MiniBooNE} results which indicate a flavor
conversion of $\bar\nu_\mu$ to $\bar\nu_e$ at the level of about
0.003. At the same time MiniBooNE has seen a low-energy excess of
events which may or may not be related to their  and
LSND's primary signal.

The results from calibrations of low-energy radiochemical solar-neutrino experiments using the reaction $\nu_e+{\rm Ga} \rightarrow {\rm Ge}
+ e^-$ based on artificial, monoenergetic neutrino sources ($^{51}$Cr
and $^{37}$Ar) show a deficit in count rate of about 25\% with
an uncertainty of about 10\%~\cite{Giunti,Mention:2011rk}.

The so-called reactor anomaly~\cite{Mention:2011rk} indicates a 6\%
deficit of $\bar\nu_e$ emitted from nuclear reactors at baselines less
than $100\,\mathrm{m}$. Interestingly, this is entirely based on the
re-analysis of existing data; the deficit is caused by three
independent effects which all tend to increase the expected neutrino
event rate. There have been two re-evaluations of reactor
antineutrino fluxes~\cite{Mueller-Huber}; both see
an increase of flux by about 3\%. The neutron lifetime decreased from
887--899\,s to 885.7\,s~\cite{PDG} and thus the inverse $\beta$-decay
cross section increased by a corresponding amount. The contribution
from long-lived isotopes to the neutrino spectrum was previously
neglected and enhances the neutrino flux at low energies.

All these hints together have a statistical significance exceeding 3$\,\sigma$ and
may be caused by one or more sterile neutrinos with a mass of roughly
$1\,\mathrm{eV}$.

There is also a somewhat ambiguous indication from cosmology of more relativistic degrees of freedom in the early Universe than the Standard Model allows, while large-scale-structure data disfavor the existence of a fourth neutrino with a mass in the eV range.

To resolve these anomalies a new series of experiments
is necessary and warranted. Several proposals exist, at both Fermilab
and CERN, to use pion decay-in-flight beams, as MiniBooNE did;
crucial differences from MiniBooNE would be use of a near detector
and, potentially, of LAr TPCs instead of scintillator detectors.
While these new proposals would constitute a significant step beyond
what MiniBooNE has done, especially in terms of systematics control,
it remains to be proven that a beam with a 1\%-level
contamination of $\nu_e$ can be used to perform a high-precision study
of a sub-percent $\nu_e$ appearance effect. In particular, 
many of these proposals involve near and far
detectors of very different sizes and/or geometrical acceptance, and
thus cancellations of systematics will be far from perfect.

The other proposed technology is a stored muon beam (``$\nu$STORM")~\cite{nustorm}. Here, the neutrinos are produced by the
purely leptonic, and therefore well understood, decay of muons,
thus the neutrino flux can be known with very high, sub-percent,
precision. The signals are wrong-sign muons which can be identified
quite easily in a magnetized-iron detector. The precise knowledge of
neutrino flux and expected very low backgrounds for the
wrong-sign muon search allow  reduction of systematics to a
negligible level, hence permitting precise measurements of the new
physics that may be behind the short-baseline anomalies. 
 
\section{Long-Baseline Neutrino Physics}

With the discovery of a large value for $\theta_{13}$, the physics
case for the next generation of long-baseline neutrino-oscillation experiments
has grown considerably stronger and one of the major uncertainties on
their expected performance has been removed. The remaining questions
are: the value of the leptonic CP phase and the quest for CP
violation; the mass hierarchy; whether $\theta_{23}$ is maximal and, if
not, whether it is larger or smaller than $\pi/4$; and of course, the
search for new physics beyond the the three-active-neutrinos paradigm.
Based on our current, incomplete understanding of the origin of
neutrino mass and the observed flavor structure in general, it is very
hard to rank these questions in relative importance, but with the
large value of $\theta_{13}$ it is feasible to design and build a
long-baseline facility which can address all three questions with high
precision and significance. 

The error on $\theta_{13}$ will keep decreasing as the reactor
measurements are refined, and Daya Bay is expected to yield a precision
which would be surpassed only by a neutrino factory. It is an
important test of the three-flavor oscillation model to see whether
the value extracted from disappearance at reactors matches that from
appearance in beams.

A combination of the existing experiments with future T2K, NO$\nu$A, and reactor
data will allow a first glimpse of the mass hierarchy, and, for favorable CP phases,  and with
extended running,  a 5\,$\sigma$
determination may be possible. Also, new atmospheric neutrino experiments
such as PINGU, ICAL at INO, and Hyper-K have, in principle, some
sensitivity to the mass hierarchy, although the actual level of significance
will  depend strongly on the achievable angular and energy resolution for
the incoming neutrino. There are also plans for a dedicated
experiment, Daya~Bay~2, which would rely not  on matter effects
but on measuring the interference of the two mass-squared
differences at a distance of about 60\,km from a nuclear reactor. It
thus seems likely that global fits will  provide a 3--5\,$\sigma$
determination of the mass hierarchy by about 2020. 
Nonetheless, a direct and precise 
test of matter effects and determination of the mass hierarchy from a single
measurement would be valuable.

A commonly used paradigm for neutrino physics beyond
oscillations is so-called non-standard interactions (NSI). These can
arise in many different models and their phenomenology is easy to
capture in a model-independent way. Large $\theta_{13}$  means that interference of
oscillation amplitudes proportional to $\sin 2 \theta_{13}$ with NSI
effects can enhance NSI sensitivity substantially. If NSI are present, the
extraction of the mass hierarchy from global fits is unlikely to
yield the correct result. Also, NSI are a straightforward
mechanism to induce a difference between  reactor and beam
measurements of $\theta_{13}$. Longer baselines generally have more
sensitivity to NSI and also allow a better separation of
oscillation from NSI.

The central physics goal of future long-baseline experiments is the
measurement of the leptonic CP phase and, potentially, the discovery of
leptonic CP violation. It is important to distinguish
these two goals: with large $\theta_{13}$ a measurement of the CP
phase to a given precision can be virtually
guaranteed; however, CP violation may or may not be present in the
lepton sector. 
We thus focus on the measurement of the CP phase since the discovery reach for CP violation directly derives from it. A determination of the CP phase requires 
measurement of any two of the following four transitions:
$\nu_e\rightarrow\nu_\mu$, $\bar\nu_e\rightarrow\bar\nu_\mu$,
$\nu_\mu\rightarrow\nu_e$, $\bar\nu_\mu\rightarrow\bar\nu_e$. However,
with long baselines, there will always be matter effects, 
which also contribute CP asymmetries; it is thus
necessary to separate this contribution from CP violation
in the mixing matrix. This separation is greatly facilitated by
exploiting $L/E$ information, ideally spanning a wide enough 
interval that more than one node of the oscillation can be
resolved. This requirement, in combination with limitations of
neutrino sources and detectors, implies that baselines
longer than 1,000\,km are needed~\cite{Diwan-etc}.
This is borne out in the LBNE
reconfiguration discussion: shorter baselines, such as those of the
existing NuMI beamline, require generally a larger exposure to reach
the same CP sensitivity. 

For ``superbeam" experiments,  systematic uncertainties will be a
major issue, since neither detection cross sections nor beam
fluxes are known to the required precision, and thus near detectors
 together with hadron-production data will play an important role.
However, these alone will not provide percent-level
systematics, since the beam at the near detector is composed mostly of
$\nu_\mu$, precluding a measurement of the $\nu_e$ cross section, but in the far detector the signals are $\nu_e$ (see  e.g.~\cite{Huber:2007em}). Unfortunately, there are no strong theory
constraints on the ratio of muon- to electron-neutrino cross sections~\cite{Day:2012gb}. To provide an independent constraint on the electron (anti)neutrino
cross section, a facility such as $\nu$STORM is crucial.

\begin{figure}[t]
\begin{center}
\includegraphics[width=0.7\textwidth]{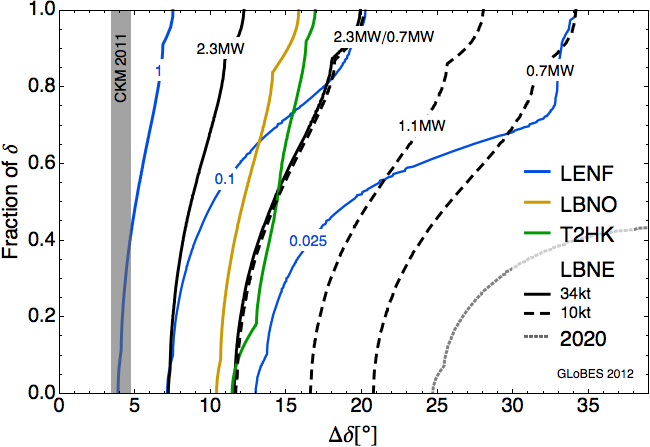}\vspace{-.2in}
\end{center}
\caption{\label{cp}  Fraction of values of the CP phase, $\delta$, for  which a given $1\,\sigma$ precision $\Delta \delta$ can be achieved, for facilities indicated in the  legend:
  LENF, a 10\,GeV neutrino factory with $1.4\times 10^{22}$ useful
  muon decays, using 4\,MW proton-beam power for
  $10^8$\,s, 2,000\,km baseline and a 100\,kt magnetized iron
  detector; LBNO, a 100\,kt LAr detector at a baseline of
  2,300\,km and $10^{22}$\,pot at 50\,GeV (about
  800\,kW of beam power for $10^8$\,s); T2HK, a 560\,kt water-Cherenkov detector at 295\,km using a 1.66\,MW beam for $5 \times
  10^7$\,s (equivalent to $1.2\times10^8$\,s at 700\,kW); LBNE,
  using LAr detectors of either 10\,kt or 34\,kt at a distance of
  1,300\,km for
  $2\times10^8$\,s with beam powers as indicated in the legend; ``2020," results from a combined fit to
  nominal runs of T2K, NO$\nu$A and Daya Bay. All detector masses are
  fiducial. The vertical gray shaded area, labeled ``CKM 2011'',
  indicates the current uncertainty on the CP phase in the CKM matrix. 
  Calculations include near detectors and assume consistent flux and
  cross-section uncertainties. (Plot courtesy P.  Coloma~\cite{Coloma:2012}.) }
\end{figure}

Figure~\ref{cp} compares CP precision for various
facilities. The 10\,GeV
neutrino factory (labeled LENF) is the only facility 
approaching the CKM precision, and has the potential even to go beyond
that. The ``superbeams" LBNE, LBNO, and T2HK, and the ``2020" global fit, span a very wide range of precision, demonstrating the crucial
importance of achieving sufficient statistics. The
number of events is determined by the product of beam power, detector
mass and running time, each of which can easily  vary
by  an order of magnitude.  LBNO has recently submitted an
expression of interest~\cite{LBNO} to CERN  outlining a much
smaller detector and lower beam power, which would put its CP precision
 close to that of any of the reconfigured LBNE options. 
 
 The sensitivity of these
results to assumptions about systematics is not shown in the figure, but important differences exist. For example, T2HK exhibits a very
strong sensitivity to the assumed level of
systematics~\cite{Coloma:2012} and thus is significantly more at risk
of being systematics-limited.  Due to
their long baselines and resultant wide $L/E$ coverage, both the LBNE Homestake option and LBNO  are quite safe from systematics~\cite{Coloma:2012}. 
At the current stage all of these experiments must rely on assumptions about
their systematics.  In any comparison such as that of  Fig.~\ref{cp},
the relative performance can vary greatly depending on these
assumptions. In the end, \emph{both} sufficient statistics and small systematics will be required in order to perform a precise
measurement of the CP phase.

It is worth noting that a neutrino factory with only 1/40th of its design
exposure has a physics performance similar to that of LBNE with 10\,kt
and 700\,kW. This low-luminosity neutrino factory could naturally evolve from
$\nu$STORM and a detector located between 1,300 and 2,300\,km
from the accelerator.  For this exposure, no muon cooling is
needed, and a proton beam of 700\,kW at 60--120\,GeV with a 10--15\,kt detector running for $2\times10^8$\,s would
suffice.  Increased beam power and the addition
of muon cooling would allow eventually reaching the full neutrino
factory exposure. A lower-energy
option with $\approx$\,5\,GeV stored-muon energy is also feasible and provides
comparable physics sensitivities. In this case, to match the lower energy, the baseline should not exceed
1,500\,km.

\section{Neutrino Physics Summary}

A staged muon-based program starting with $\nu$STORM, which can evolve
in various, adjustable steps to a full Neutrino Factory and pave the way towards a Muon Collider, seems a
very attractive option. It will produce outstanding physics at each stage.  

\section{Collider Physics Landscape}

The Standard Model (SM) has been a spectacular success.  For more than thirty years, all new observations except for neutrino mass have fit naturally into this framework. But basic questions 
remain: (1) Is the Higgs mechanism the correct mechanism of electroweak symmetry
breaking?  (2) How do the fermion masses and flavor mixings arise?  Furthermore, 
the Standard Model is incomplete. It does not explain dark matter; neutrino masses 
and mixings require new particles or interactions; and the observed baryon asymmetry 
in the universe  requires additional sources of CP violation.  From a theoretical 
viewpoint there are also problems with the SM.  It has been argued by G. 't~Hooft 
that the SM is not natural at any energy scale $\mu$ much above the Terascale ($\sim$\,1 TeV) 
because the small dimensionless parameter $\epsilon(\mu) = (m_H/\mu)$ is not 
associated with any symmetry as $\epsilon \rightarrow 0$ \cite{Hooft:1979bh}.  
This is the naturalness problem of the SM.  If the SM is valid all the way up to the
Planck scale ($\Lambda_{Pl} \sim 10^{19}$\,GeV), then the SM has to be fine-tuned
to a precision of one part in $(m_H/\Lambda_{PL})^{-2}$!  In this decade, the physics 
of the Terascale will be explored at the LHC.  Planned experiments studying neutrino 
oscillations, quark/lepton flavor physics, and rare processes may also provide insight 
into new physics at the Terascale and beyond.
   
Discoveries made at the LHC will elucidate the origin of electroweak symmetry breaking.  
Is that mechanism just the SM Higgs scalars or does it involve new physics? New physics 
might be new gauge bosons, additional fermion generations or fundamental scalars.  
It might be SUSY or new dynamics or even extra dimensions.  Significant theoretical 
questions will likely remain even after the full exploitation of the LHC---most notably, 
the origin of fermion (quark and lepton) masses, mixings and CP violation; the character 
of dark matter; and detailed questions about spectrum, dynamics, and symmetries of any 
observed new physics.    Thus, it is hard to imagine a scenario in which a multi-TeV 
lepton collider would not be required to fully explore the new physics.
   
To prepare for the energy frontier in the post-LHC era, research and development is 
being pursued on a variety of lepton colliders:  a low-energy ($E_{cm} < 1$\,TeV) linear 
electron-positron collider (ILC), a second design (CLIC) capable of higher energies 
($E_{cm} = 3$\,TeV), and a multi-TeV Muon Collider. 
   
A multi-TeV Muon Collider provides a very attractive possibility for studying the 
details of Terascale physics after the LHC.  Physics and detector studies 
are under way to understand the required  collider parameters (in particular 
luminosity and energy) and map out, as a function of these parameters, the associated 
physics potential.  The physics studies will set benchmarks for various new-physics 
scenarios (e.g., SUSY, Extra Dimensions, New Strong Dynamics) as well as Standard Model 
processes.  
   
Furthermore, the Muon Collider  lends itself to a staged program with 
physics at each stage of producing and cooling the muons.  Important physics 
opportunities include the possibility of a Higgs Factory and/or a Neutrino Factory 
as steps to a Muon Collider. 

\section{Muon Collider Physics}

The Muon Collider is an energy-frontier machine.  It offers both discovery as well 
as precision measurement capabilities. The physics goals of a Muon Collider (MC) 
are for the most part the same as those of a linear electron-positron collider  at the same energy (CLIC)
\cite{Accomando:2004sz, 2012arXiv1202.5940L}.  The main advantages 
of a MC are the ability to study the direct ($s$-channel) production of scalar 
resonances, a much better energy resolution (because of the lack of significant 
beamstrahlung), and the possibility of extending operations to very high energies. 
(Ideas for $E_{cm} \ge 6$\,TeV are being  entertained.)
At CLIC, however, significantly greater polarization of the initial beams is 
possible \cite{2012arXiv1202.5940L}. Furthermore, CLIC is free of the significant detector backgrounds 
found in a MC  due to muon decays upstream of the interaction point.
 
There are basically three kinds of channels of interest for a lepton collider: 
(1) open pair production, (2) $s$-channel resonance production, and (3) fusion processes.

\subsection{Pair Production}

The kinematic thresholds for pair production of standard model particles ($X + \bar{X}^{\prime}$) 
are well below $E_{cm} = 500$\,GeV.  For standard model particles at 
$E_{cm}  > 1$\,TeV  the typical open-channel pair-production process is well above its kinematic threshold and $R$ (the cross section normalized to that for annihilation into electron-positron pairs) becomes nearly flat.

For the MC a forward/backward angular cut (e.g., $10^{\circ}$) is imposed on the outgoing pair. 
Closer to the beam direction, a shielding wedge is 
needed to suppress detector backgrounds arising from the effects of muon decay in the beams.

For a process whose rate is one unit of $R$, an integrated luminosity of $100$\,fb$^{-1}$ at 
$E_{cm} = 3$\,TeV yields ~1000 events.  For example, the rate of top-quark pair 
production at 3\,TeV is only 1.86 units of $R$.  This clearly demonstrates the need for high 
luminosity in a multi-TeV lepton collider.

\subsection{Resonances}

Many models beyond the SM predict resonances that may be produced directly in the $s$ channel 
at a Muon Collider. Here, the narrow beam energy spread of a Muon Collider, $\delta E/E \sim {\rm ~few~}\times 10^{-5}$, could be an important advantage.  The cross section for the production of an $s$-channel resonance, $X$, with spin $J$, mass $M$ and width $\Gamma$ is given by
\begin{equation}
\sigma(\mu^+\mu^-\rightarrow X) = \frac{\pi}{4k^2}(2J+1)\frac{\gamma^2}{(E-m)^2 + \Gamma^2/4} B_{\mu^+\mu^-} B_{\rm visible}\,,
\end{equation}
where $k$ is the momentum of the incoming muon and $E$ the total energy of the initial system 
($E_{cm}$).  $B_{\mu^+\mu^-} \Gamma$ is the partial width of $X \rightarrow \mu^+\mu^-$ and $B_{\rm visible}$ is the visible decay width 
of $X$.  At the peak of the resonance with negligible beam energy spread, we have
\begin{equation}
R_{\rm peak}  = 3 (2J+1) \frac{B_{\mu^+\mu^-} B_{\rm visible}}{\alpha_{\rm EM}^2}\,.
\end{equation}

\subsubsection{$Z'$}

For a sequential standard model $Z^\prime$ gauge boson, the value of $R_{\rm peak}$ is strikingly large, 
typically $\sim 10^4$.  The luminosity, $L$, for $1.5 < M(Z') < 5.0$\,TeV required to produce 
1000 events on the $Z'$ peak is only 0.5--$5.0 \times 10^{30} \,{\rm cm}^{-2}\,{\rm s}^{-1}$.  Hence, a comprehensive first-order study of the properties of a narrow resonance, such as a $Z'$, in the few-TeV 
mass range can be carried out with a low-luminosity, $L \sim 10^{30}\,{\rm cm}^{-2}\,{\rm s}^{-1}$, Muon Collider.

\subsubsection{Extra Dimensions}

Theories with extra dimensions that have a radius of curvature close to the 
Terascale have been postulated.  Here one expects an excitation spectrum, arising from excited modes in 
the extra dimension, for the graviton and any other SM particles whose interactions occur in the bulk. 
In the Randall-Sundrum warped extra dimensions models~\cite{Randall}, the graviton 
spectrum contains additional resonances (KK-modes) that can be probed by a Muon Collider.

From the perspective of energy frontier colliders, however, only the physics at the 
first (perhaps second) Kaluza-Klein (KK) mode will be relevant.  All kinematically allowed 
KK-mode resonances are accessible to a multi-TeV Muon Collider. These include the $Z'$ 
and $\gamma'$ of the electroweak sector.  The precise measurement of the $Z'$ and $\gamma'$ mass scales will determine the various electroweak symmetry-breaking structures, and how these states 
couple to different fermion generations will determine bulk fermion localization.  

\subsection{Fusion Processes}

\begin{figure}[t]
\vspace{-.4cm}
\begin{minipage}[b]{0.5\linewidth}
\centering
~~~~~~~~~~~~~\includegraphics[width=0.75\textwidth, trim=280 100 200 100, clip]{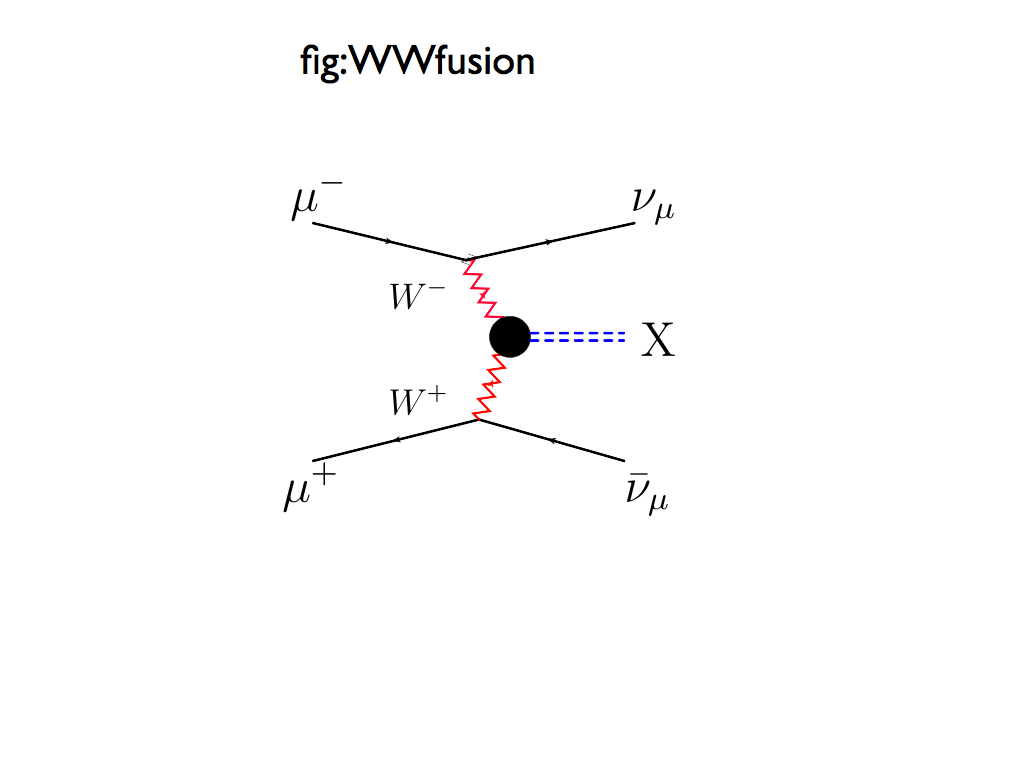}
\vspace{-.5cm}
\caption{\label{fig:WW}{Typical fusion process.}}
\end{minipage}
\begin{minipage}[b]{0.48\linewidth}
\centering
\hspace{0.5cm}\includegraphics[width=0.75\textwidth, trim=280 100 260 150, clip]{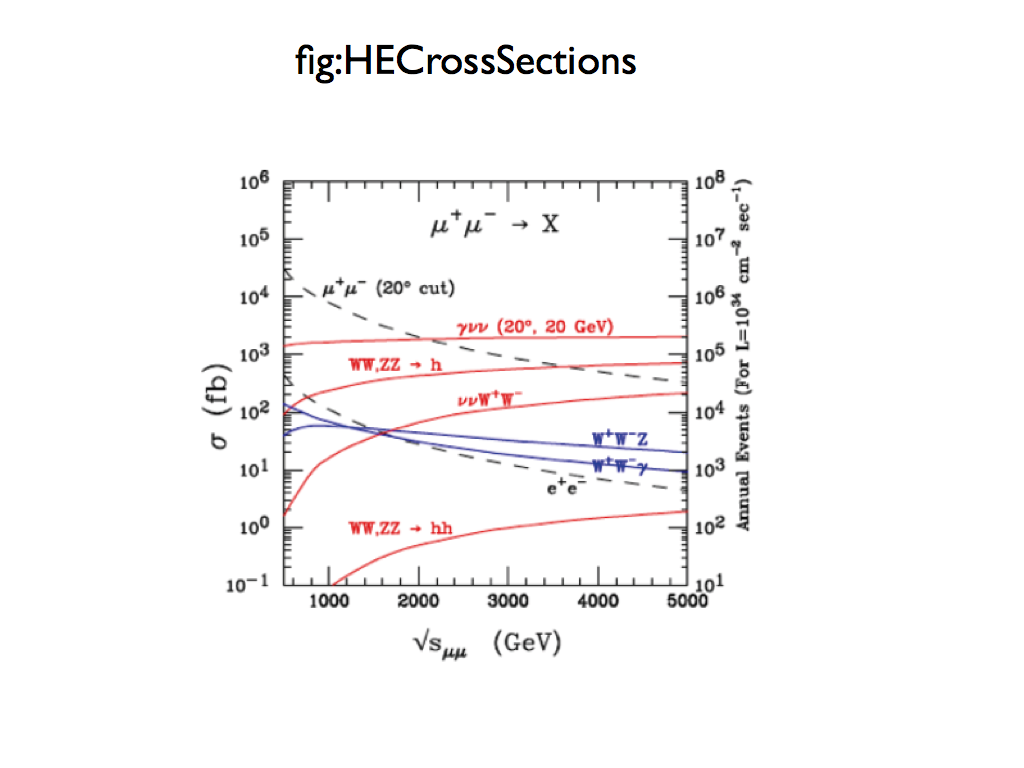}
\vspace{-.2cm}
\caption{\label{fig:HE}{Various fusion cross sections versus $\sqrt{s}$ at lepton colliders.}}
\end{minipage}
\end{figure}

A typical fusion process is shown in Fig.\ \ref{fig:WW}.  For $E_{cm} >\!\!> M_X$ 
the cross section is typically large and grows logarithmically with $s=E_{cm}^2$; while the usual 
pair-production processes are constant in $R$ and thus dropping like $1/s$.  Thus, for asymptotically 
high energies, fusion processes dominate.  For lepton colliders, the crossover occurs 
in the few-TeV region in standard model processes, as shown in Fig.\ \ref{fig:HE}. A variety of processes 
are shown including $WW$ and $ZZ$ inclusive production.  The large rates for $WW,$ $WZ$, and $ZZ$ fusion processes imply that the multi-TeV Muon Collider is also effectively an electroweak-boson 
collider.

Physics studies of fusion processes such as $\mu^+ \mu^- \rightarrow Z^* Z^* \mu^+ \mu^- \rightarrow  X \mu^+ \mu^-$ benefit greatly 
by tagging of the outgoing $\mu^{\pm}$ and hence will be sensitive to the required $\approx10^{\circ}$ angular cut.

\section{Standard Model Higgs Boson}

Both the ATLAS \cite{Aab:2012an} and CMS \cite{Chatrchyan:2012tx} experiments at the LHC have observed a new boson 
with a mass about 126 GeV.   Further data will be required to establish the spin ($J=0 {\rm ~or~} 2$),  parity, and determine whether the branching
ratios to various final states are consistent with the Standard Model Higgs boson.  But whatever the results, it is clear that more precise
studies of this new boson will be an important target of future lepton colliders.  A Muon Collider Higgs Factory has the unique ability to allow the direct measurement of the width of the Higgs-like boson.

Studies of the feasibility of direct production of the SM Higgs boson were carried out 
over a decade ago \cite{Barger:1996pv} for a low-energy, high-luminosity MC.  It was found that 
very precise control of the beam energy and energy spread are required. The discovery of a
new Higgs-like boson at the LHC has renewed interest in a Muon Collider Higgs Factory 
\cite{Neuffer:2012zz}.

Higgs bosons can be studied in a number of other ways at a multi-TeV Muon Collider:

\begin{enumerate}

\item Associated production: $\mu^+ \mu^- \rightarrow Z^* \rightarrow Z^0 + h^0$  has $R \sim0.12$.  We can measure the $b$-quark Higgs-Yukawa coupling and look for invisible decay modes of the Higgs boson.

\item Higgsstrahlung: $\mu^+ \mu^- \rightarrow t \bar{t} + h^0$  has $R \sim 0.01$ (so such a study requires $\sim 5\,{\rm ab}^{-1}$).  This could provide a direct measurement of the top quark Higgs-Yukawa coupling.

\item $W^*W^*$ fusion into $\bar{\mu}_{\mu} \nu_{\mu} + h^0$ has $R \sim 1.1\, s \ln{s}$ (for $m_h = 120$\,GeV).  It allows the study of Higgs self-coupling and certain rare decay modes.

\end{enumerate}

\section{Extended Higgs Sector}

\begin{figure}[t]
\vspace{-.5cm}
\begin{center}
\hspace{1cm}\includegraphics[width=0.7\textwidth, trim=100 160 200 120, clip]{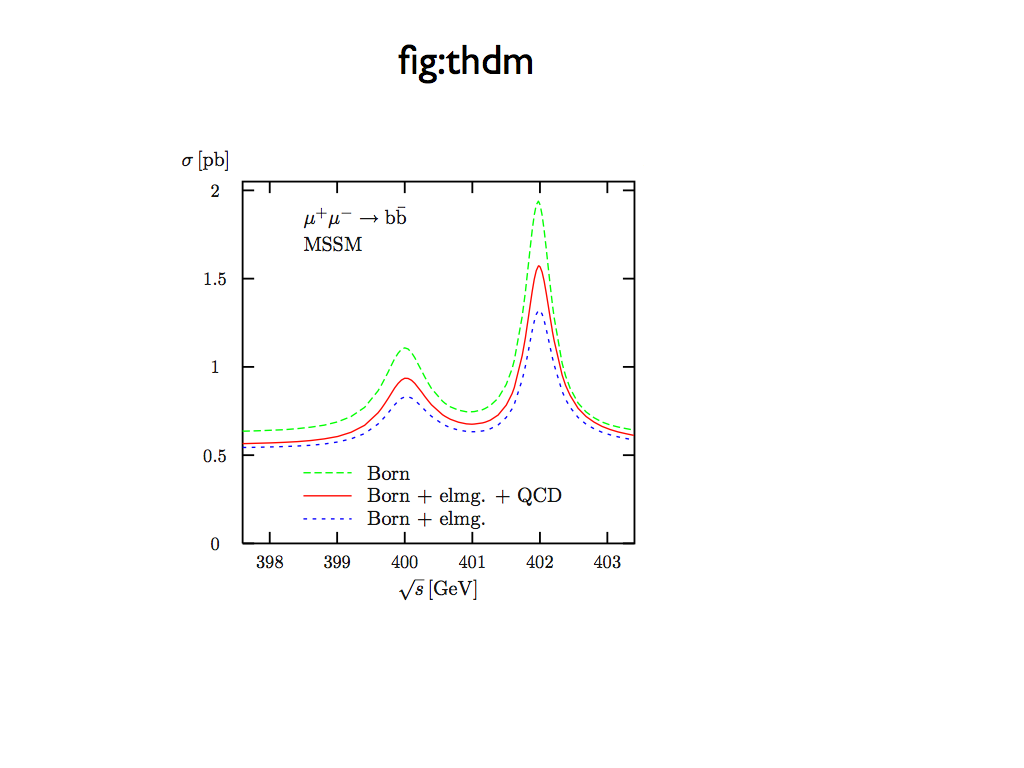}
\end{center}
\vspace{-0.7cm}
\caption{\label{2hdm}{MSSM cross section $\mu^+ \mu^- \rightarrow b \bar{b}$  near the $H^0$ and $A$ resonances (with $M_A = 400$\,GeV 
and $\tan \beta = 5$).  SM radiative corrections to the lowest-order processes have been included. (From \cite{Dittmaier:2002nd}.)}}
\end{figure}

In the two-Higgs doublet scenario there are five scalars: two charged scalars $H^\pm$, two 
neutral CP-even scalars $h$, $H^0$,  and a neutral CP-odd scalar $A$.  
Two Higgs doublets occur naturally in supersymmetric models \cite{Djouadi:2005gj}. 
For the constrained MSSM model (cMSSM), 
as the mass of the $A$ is increased, the $h$ becomes closer to the SM Higgs couplings and the other 
four Higgs become nearly degenerate in mass. This makes resolving the two 
neutral-CP states difficult without the good energy resolution of a Muon Collider, as shown in Fig.\ \ref{2hdm}.

\section{Supersymmetry}

Supersymmetry (SUSY) provides a solution to the naturalness problem of the SM.  It is a 
symmetry that connects scalars with fermions, ordinary particles with superpartners:
a symmetry that is missing in the SM.  

The simplest SUSY model is the constrained MSSM, with only five parameters determining the masses 
of all the superpartners. cMSSM scenarios are now highly constrained  by the direct observation of a new Higgs-like boson 
at a mass of 126 GeV \cite{Arbey:2011ab}.  Powerful constraints on cMSSM are also provided by the electroweak 
precision measurements and the lack of any indications  of SUSY from flavor physics so far 
\cite{Buchmueller-etc}.  

Arbey {\it et al.}\cite{Arbey:2011ab}  found that cMSSM models 
with minimal anomaly- or gauge-mediated supersymmetry breaking are now disfavored as these models predict a too light Higgs particle.
The gravity-mediated constrained MSSM would still be viable, provided the scalar top quarks are heavy and their trilinear coupling large. Significant areas of the parameter space of models with heavy supersymmetric particles, such as split or
high-scale supersymmetry, could also be excluded as, in turn, they generally predict a
too heavy Higgs particle\cite{Arbey:2011ab}.

To date, no evidence for SUSY particles has been found at the LHC.  In particular, ATLAS \cite{Aad:2011ib} and CMS \cite{Chatrchyan:2012mf} have produced strong direct bounds on the masses of squarks and gluinos below 1\,TeV.

All this, taken together, makes it almost certain that direct coverage of the remaining 
MSSM parameter space requires a multi-TeV--scale lepton collider such as CLIC or 
a Muon Collider.

\section{New Strong Dynamics}

Strong dynamical models of electroweak symmetry breaking have no elementary scalars and 
thus avoid the naturalness problem of the SM.  Chiral symmetry breaking (\`a la QCD) in the 
technicolor sector produces technipions that give the proper masses to the $W$ and $Z$ bosons.  
For details and a discussion of various new strong dynamics models see the review of Hill 
and Simmons\cite{Hill:2002ap}.

The minimal Technicolor model contains an isospin triplet techni-rho ($\rho_T$) and singlet 
techni-omega ($\omega_T$) vector mesons, which can be produced in the $s$ channel in lepton colliders. 
In addition, it contains a techni-eta$^\prime$ ($\eta_T'$) which would be produced in association with 
$Z$ bosons in analogy to the Higgs boson.

\begin{figure}[ht]
\vspace{-.2cm}
\begin{minipage}[b]{0.5\linewidth}
\centering
\includegraphics[width=0.99\textwidth, trim=170 100 150 100, clip]{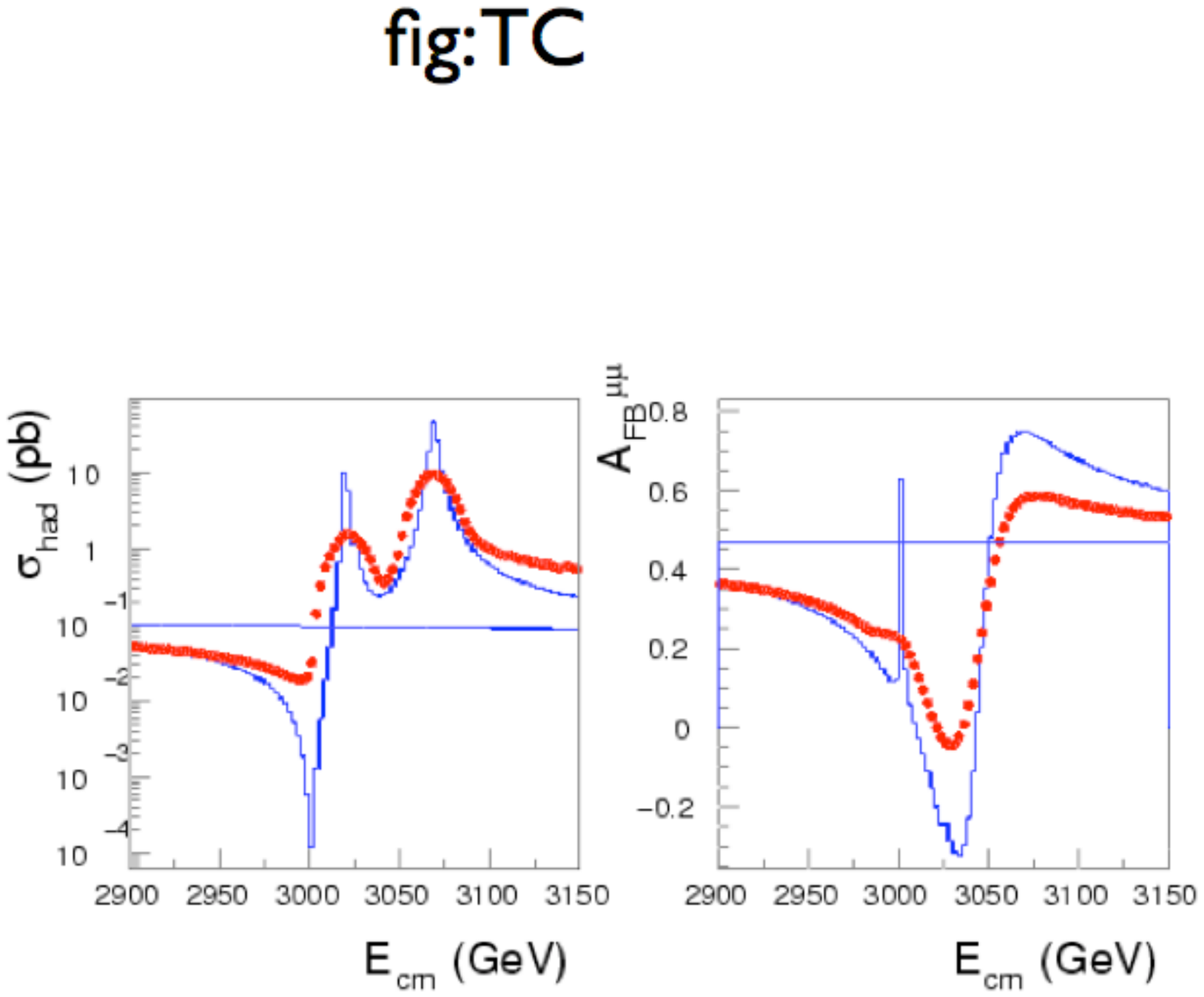}
\vspace{-1cm}
\caption{\label{TC}{D-BESS model at CLIC \cite{Anichini:1994xx, Casalbuoni:1995qt}. CLIC energy resolution limits ability to disentangle nearby states 
expected in models with new strong dynamics.}}
\end{minipage}
\hspace{0.5cm}
\begin{minipage}[b]{0.45\linewidth}
\centering
\includegraphics[width=0.7\textwidth]{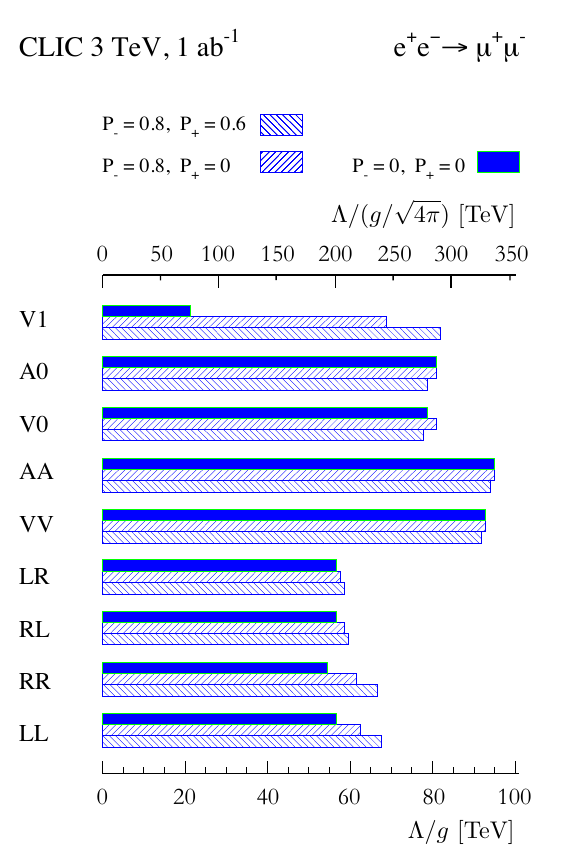}
\vspace{-.2cm}
\caption{\label{fig:contact}{Reach of a lepton collider for contact terms with various chiral structure.  The role of polarization is 
indicated. (From \cite{2012arXiv1202.5940L}.)}}
\end{minipage}
\end{figure}

In less minimal schemes, there are residual techni-pions, $\pi_T^{\pm}$ and $\pi_T^0$, that can be produced in lepton colliders. The techni-rho is typically broad if the two--techni-pion channel is 
open but, as in QCD, the techni-omega is nearly degenerate and narrow.  In low-scale 
Technicolor models, some techni-rho ($\rho_T$) can be light ($\sim$\,250 GeV) as well as nearly 
degenerate in mass with a techni-omega, and these can be studied in great detail at a 
Muon Collider with the appropriate energy~\cite{Hill:2002ap}.  For techni-rho and techni-omega masses 
in the TeV range, a CLIC study has been done to determine its resolving power.  The results 
for a Muon Collider are essentially the same as the CLIC curve before inclusion of 
beamstrahlung and ISR effects.  For this physics, the Muon Collider has a distinct advantage 
over CLIC. The comparison is shown in Fig.\ \ref{TC}. 

There are other approaches to new strong dynamics:  Topcolor, TC2, composite and Little Higgs models
\cite{Hill:2002ap}.  All of these would provide a rich spectrum of states that can be 
observed at a multi-TeV Muon Collider.

\section{Contact Interactions}

New physics can enter through contact interactions, which are higher-dimension operators 
in the effective Lagrangian, as
\begin{equation}
{\cal L} = \frac{g^2 \bar{\psi}\Gamma {\psi} \bar{\psi} \Gamma {\psi}}{\Lambda^2}\,.
\end{equation}

The MC is sensitive to $\Lambda \sim 200$\,TeV, roughly equivalent to CLIC.  Preliminary studies 
suggest that the forward-angle cut is not an issue here \cite{Eichten:1998kn}.
If polarization is not available at a MC, it may be at a disadvantage compared with CLIC 
in the ability to disentangle the chiral structures of the new operators (see Fig.\ \ref{fig:contact}).

\section{Muon Collider Physics Summary}
  
A multi-TeV Muon Collider is required for the full coverage of Terascale physics.  
The physics potential for a Muon Collider at $\sim3$\,TeV and integrated luminosity of $1\,{\rm ab}^{-1}$ 
is outstanding.  Particularly strong cases can be made if the new physics is SUSY or new 
strong dynamics. Furthermore, a staged muon collider can provide a Neutrino Factory to 
fully disentangle neutrino physics. If a narrow $s$-channel resonance state exists in 
the multi-TeV region, the physics program at a Muon Collider could begin with less than 
$10^{31}\,{\rm cm}^{-2}\,{\rm s}^{-1}$ luminosity.  

The observation of a new state at 126 GeV by both ATLAS and CMS revitalizes  consideration 
of a Higgs factory as part of a staged multi-TeV muon collider.  This is particularly attractive if there is an enlarged scalar sector (e.g., THDM, SUSY). Many details will remain to be understood even after the LHC.  

Detailed studies of the physics case for a  multi-TeV muon collider are just beginning.  
The goals of such studies are to: (1) identify benchmark physics processes; (2) study the 
physics dependence on beam parameters; (3) estimate detector backgrounds; and (4) compare 
the physics potential of a Muon Collider with those of the ILC, CLIC and upgrades to the LHC.

\section{A Staged Muon Accelerator Program}

To elaborate on points already mentioned, advantages of muon over electron colliders are:
\begin{enumerate}
\item Synchrotron radiation, proportional to $E^4/m^4$,  is strongly suppressed for muons, allowing circular machines much smaller than an $e^+e^-$ collider of  comparable energy.
\item Each muon bunch thus collides repeatedly, allowing larger emittances and fewer leptons for a given luminosity. The number of collisions per machine cycle is limited by the muon lifetime to $\approx$\,150 times the average bending field (typically 7\,T), corresponding to $\sim$\,1000 turns, and thus this number of bunch crossings in each detector. In contrast, in an electron-positron linear collider the beams interact only once.
\item In a circular collider ring, there can be more than one detector (two in our design), which effectively doubles the luminosity.
\item Radiation during bunch crossings (beamstrahlung) is also proportional to $E^4/m^4$, giving a much smaller energy spread at a Muon Collider than at an $e^+e^-$ collider.
\item $s$-channel Higgs production is enhanced by the factor $(m_\mu/m_e)^2\approx 40,000$, making exploitation of such production practicable, whereas with electrons it is not.
\end{enumerate}

Muon accelerators require  technologies already in use, or that are straightforward extensions of the current state of the art: linacs and recirculating linear accelerators (RLAs) are already fast enough for muon acceleration. The focus of muon-acceleration R\&D is thus cost reduction, which can be achieved by means of newly developed fixed-field alternating gradient accelerators (FFAGs) and very-rapid-cycling synchrotrons.

Achieving the $\mu$m-scale beam emittances desired for high luminosity is more challenging, but feasible via ionization cooling~\cite{cooling}. In this, the muons are alternately passed through energy-absorbing media and reaccelerated by means of RF cavities, while being strongly focused by superconducting solenoids in order to overcome the heating effects of multiple Coulomb scattering. The evolution with path-length $s$ of the normalized transverse emittance $\epsilon_n$ of muons of energy $E_\mu$, mass $m_\mu$, and speed $\beta$ traversing a material medium is given by~\cite{Neuffer-Fernow}
\begin{equation}
\frac{d\epsilon_n}{ds}\ \approx\
-\frac{1}{\beta^2} \left\langle\frac{dE_{\mu}}{ds}\right\rangle\frac{\epsilon_n}{E_{\mu}}\ +
\ \frac{1}{\beta^3} \frac{\beta_\perp (0.014\,{\rm GeV})^2}{2E_{\mu}m_{\mu}X_0}\,,
\label{eq:cool}
 \end{equation} 
where $\beta_\perp$ is the lattice focal length and $X_0$ the radiation length of the medium. The large $dE/ds$ (4.0\,MeV/(g/cm$^2$)) and long radiation length (61.3\,g/cm$^2$) of hydrogen make it, and such compounds as LiH, the media of choice, while solenoids with fields up to 30\,T are in development. Together these allow transverse emittances down to $\sim$\,10\,$\mu$m to be achieved---a reduction factor of $10^6$ in six-dimensional emittance---enabling  $\sim10^{34}\,{\rm cm}^{-2}\,{\rm s}^{-1}$ luminosity. The MICE experiment~\cite{MICE-JINST} in progress at the UK's Rutherford Appleton Laboratory will demonstrate muon ionization cooling and, starting in 2013, provide measurements needed for detailed validation of the simulation codes used to design ionization cooling channels.

The remaining ingredients for colliders of such luminosity are multi-megawatt proton beams and targets; these have been developed for spallation neutron sources, and the MERIT experiment at CERN has established the feasibility of the configuration of choice for muon facilities: a free-flowing mercury jet in a high-field ($\approx$\,15\,T) hybrid copper-insert/ super\-conducting-outsert pion-capture solenoid. The ``Project X" accelerator proposed at Fermilab is suitable and can be upgraded to the  beam power that will ultimately be needed.

The same technologies, albeit in less demanding incarnations, enable Neutrino Factories producing beams of $\sim$\,$10^{21}$ electron and muon neutrinos per year aimed at remote detectors over $\sim$\,1,000\,km baselines. In particular, a factor $\sim$\,10 in emittance reduction suffices for such a high-luminosity Neutrino Factory, and, as mentioned, an entry-level facility without cooling, to which cooling could be added as an upgrade, is already of considerable interest. Adding a true NF front-end to $\nu$STORM could increase the flux a factor of 10--50, and
the decay ring magnets could then be reused in a new tunnel oriented on a longer baseline.

The International Design Study for a Neutrino Factory (IDS-NF) is targeting a Reference Design Report on the 2013 timescale. In the US, the Muon Accelerator Program (MAP) is in progress with the goal of completing the R\&D necessary to validate the Muon Collider concept on a 6-year timescale.

\begin{figure}
\centerline{
\includegraphics[width=1.65in]{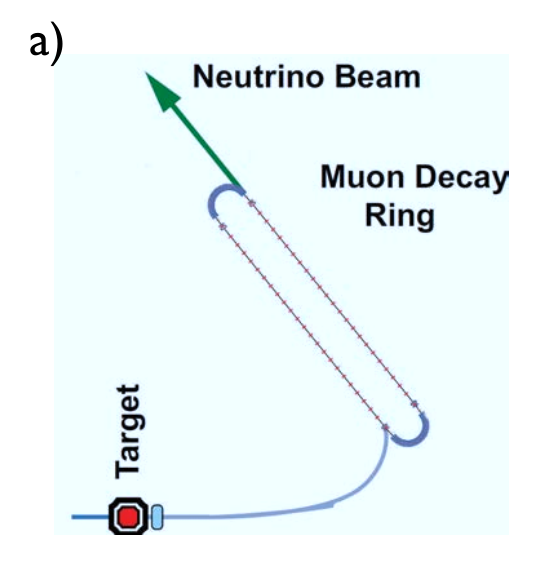}
\includegraphics[width=3.8in]{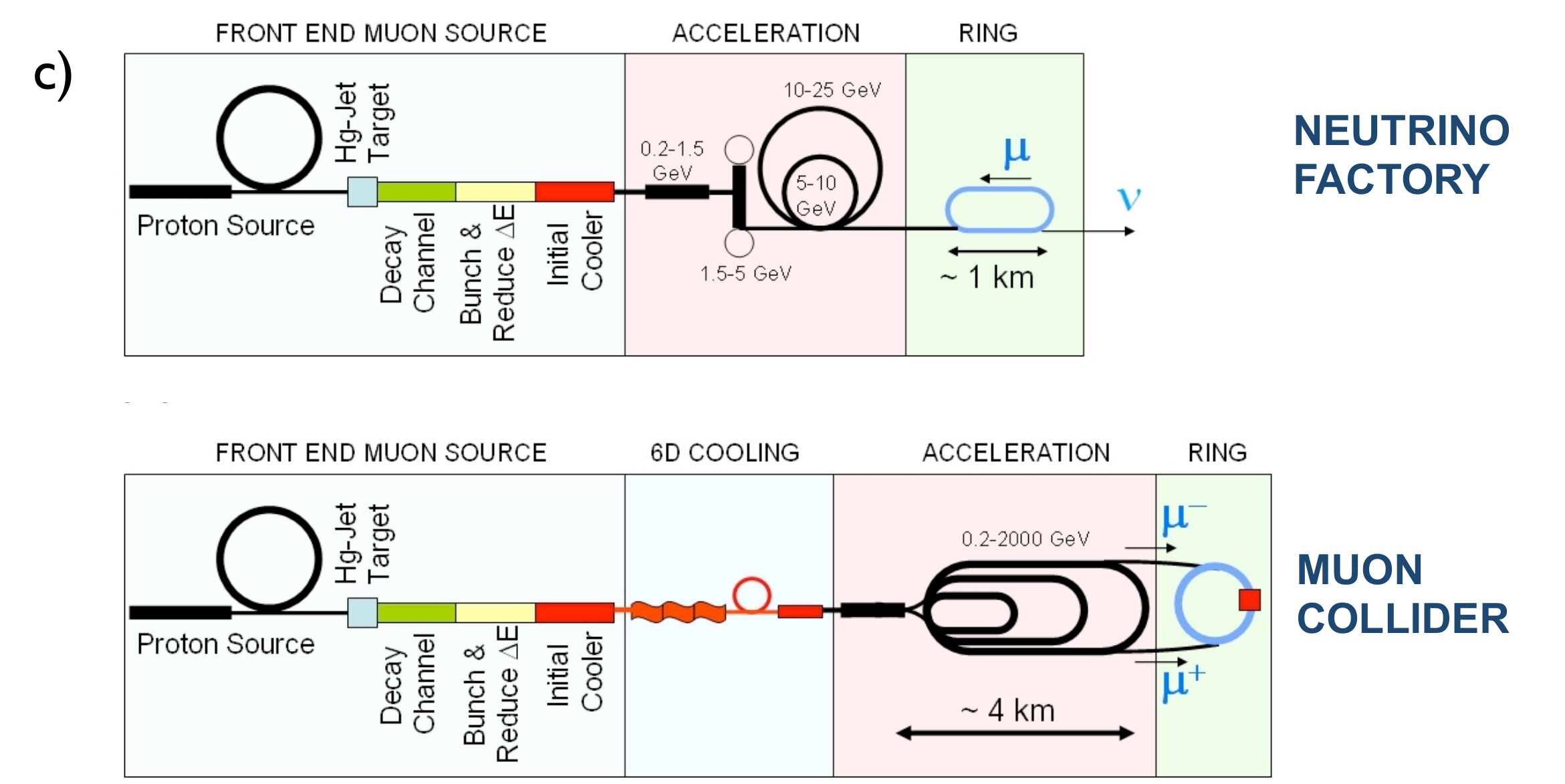}}
\centerline{
\includegraphics[width=2.5in]{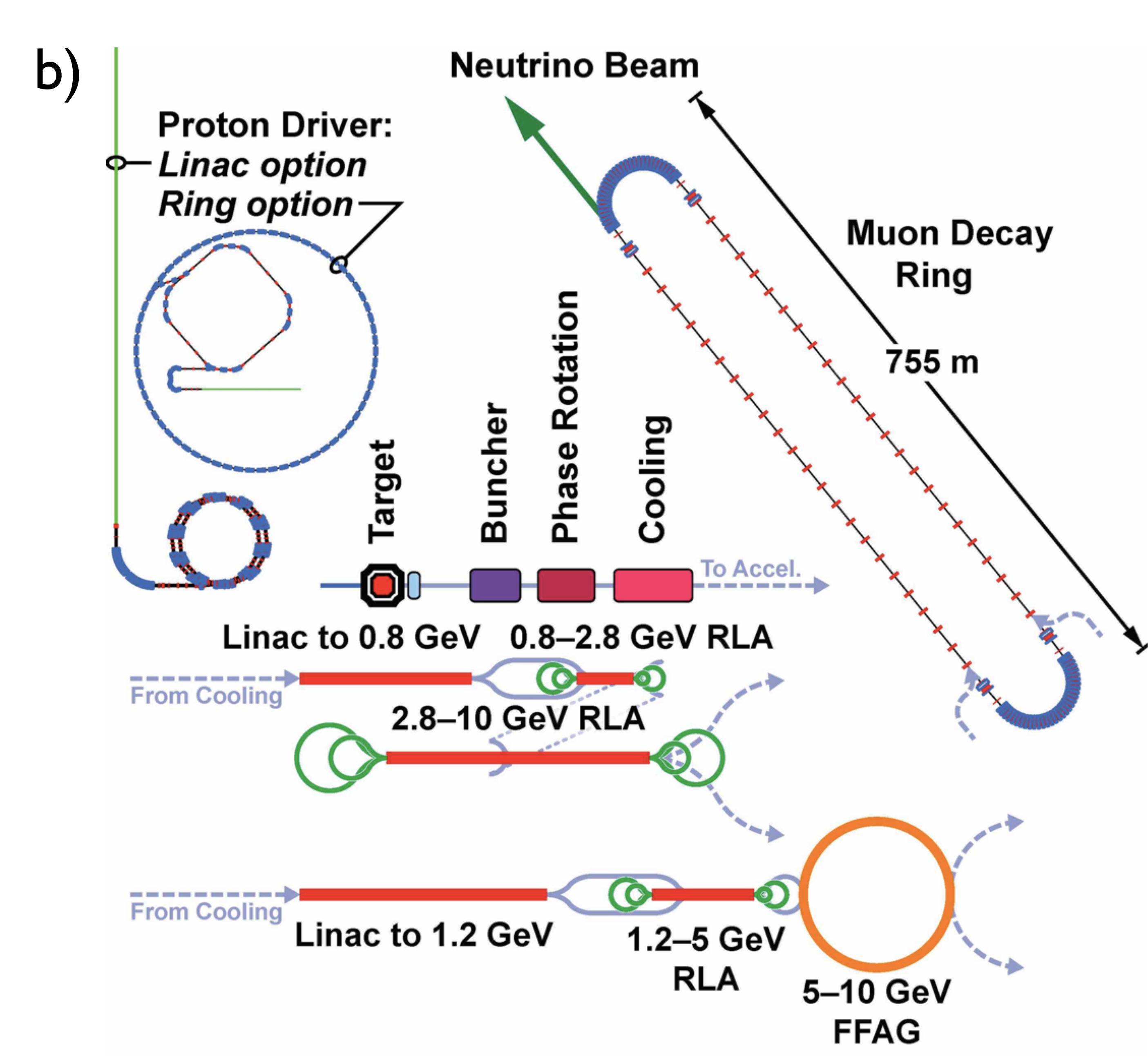}
\includegraphics[width=3in,trim=3 0 3 0 mm,clip]{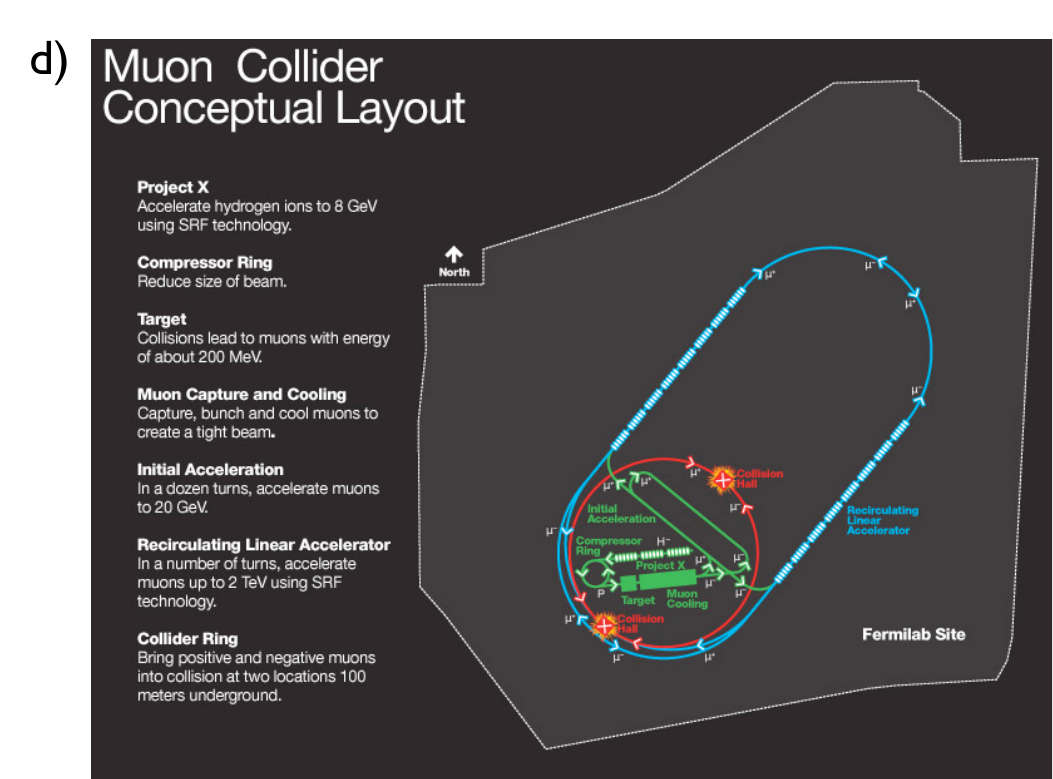}}
\caption{Ingredients for a possible muon-facility staging plan: a) $\nu$STORM; b) 
IDS-NF Neutrino Factory; c) Muon Collider as Neutrino Factory upgrade; d) possible siting of multi-TeV Muon Collider at Fermilab.}\label{fig:staging}
\end{figure}

Ingredients for a possible staging sequence are illustrated in Fig.~\ref{fig:staging}. Since the LHC boson discovery and measurement of $\theta_{13}$ are so recent, whether a Neutrino Factory should be a step on the way to a Higgs Factory or vice versa as yet remains to be determined.
However, decisions on the most effective route for exploring these frontiers are rapidly approaching. Renewed world attention to these opportunities is thus a matter of pressing interest.

\section*{Acknowledgments}
We thank Patrick Huber and Estia Eichten for extensive and valuable input to this report.

\bibliographystyle{aipprocl}

\bibliography{references}

\end{document}